\documentclass[%
twocolumn,
 xcolor,
 amsmath,amssymb,esint,commath,
 aps,
 prd,
]{revtex4-1}

\usepackage{graphicx}  
\usepackage{dcolumn}   
\usepackage{bm}        
\usepackage{amssymb}   
\usepackage{hyperref}  
\usepackage[mathlines]{lineno}  
\usepackage{mathrsfs} 
\usepackage{float}
\usepackage{comment}  

\usepackage{color}
\usepackage{booktabs} 
\usepackage{xspace}

\newcommand{\Finesse}{\textsc{Finesse}\xspace}
\newcommand{\finesse}{\mathscr{F}}
\newcommand{\HG}[1]{$\text{HG}_{\text{#1}}$}
\newcommand{\FSR}{\text{FSR}}

\newcommand{\VperrootHz}{$\text{V}_{\text{rms}}/\sqrt{\text{Hz}}$ $ $}

\definecolor{linkcolor}{rgb}{.8,0,0}
\definecolor{urlcolor}{rgb}{0,0,.7}
\definecolor{citecolor}{rgb}{0,.5,0}
\definecolor{acrocolor}{rgb}{0,0,.7}
\hypersetup{bookmarksopen,colorlinks=true}
\hypersetup{pdfstartview=FitH}
\hypersetup{linktocpage=true,bookmarksnumbered=true}
\hypersetup{plainpages=false,breaklinks=true}
\hypersetup{linkcolor=linkcolor,citecolor=citecolor,urlcolor=urlcolor}

\begin{document}


\title{Feasibility of near-unstable cavities for future gravitational wave detectors}
\author{Haoyu Wang$^{1,2}$}
\email{haoyu@star.sr.bham.ac.uk}
\author{Miguel Dovale-\'Alvarez$^1$}
\author{Christopher Collins$^1$}
\author{Daniel David Brown$^1$}
\author{Mengyao Wang$^1$}
\author{Conor M. Mow-Lowry$^1$}
\author{Sen Han$^2$}
\author{Andreas Freise$^1$}
\affiliation{$^1$ School of Physics and Astronomy and Institute of Gravitational Wave Astronomy,
University of Birmingham, Edgbaston, Birmingham B15 2TT, United Kingdom}
\affiliation{$^2$ Lab of Optical Instrument and Precision Measurement, School of Optical Electrical and Computer Engineering, University of Shanghai for Scienceand Technology, Shanghai 200093, China}

\date{\today}

\begin{abstract}
Near-unstable cavities have been proposed as an enabling technology for future gravitational wave detectors, as their compact structure and large beam spots can reduce the coating thermal noise of the interferometer. We present a tabletop experiment investigating the behaviour of an optical cavity as it is parametrically pushed to geometrical instability. We report on the observed degeneracies of the cavity's eigenmodes as the cavity becomes unstable and the resonance conditions become hyper-sensitive to mirror surface imperfections. A simple model of the cavity and precise measurements of the resonant frequencies allow us to characterize the stability of the cavity and give an estimate of the mirror astigmatism. The significance of these results for gravitational wave detectors is discussed, and avenues for further research are suggested.
\end{abstract}

\maketitle

\section{Introduction}\label{sec:Introduction}

Mirror coating thermal noise is a major source of noise in advanced gravitational wave detectors over much of the signal band~\cite{harry2002}. One way to reduce this source of noise is to use a cavity configuration with a larger beam spot size on the mirrors~\cite{harry2012}. This requires, however, that the cavity parameters get close to the boundary of geometric stability~\cite{barr2012}. Fabry-Perot cavities operating close to the edge of geometrical instability may be driven into the unstable region via small cavity length perturbations or mirror surface distortions. We refer to these devices as near-unstable cavities (NUC), and as such, NUCs are at risk of suffering from the consequent problems of high optical loss and Gaussian mode degeneracy~\cite{Siegman67, gretarsson07}. The well-defined Gaussian beams in NUCs are also prone to become distorted through their interaction with the small imperfections of the mirror surfaces. These issues have an adverse impact on the detector sensitivity and controllability.



We report on the design and operation of a tabletop experiment to investigate the behaviour of NUCs. A cavity was built and accurate control achieved through length and alignment sensing systems. The cavity was then parametrically pushed into the geometrically near-unstable and unstable regions. We provide a detailed account of the design challenges and technical hurdles associated with this type of cavity, as well as their behaviour as they gradually become unstable. Our results provide an insight into how far the cavity parameters can be pushed towards instability whilst maintaining controllability and spatial filtering. The work presented will aid the prototype experiment~\cite{graf2012} and the design of future ground-based gravitational wave detectors.



\begin{figure}[!ht]
\begin{center}
\includegraphics[width=0.99\linewidth]{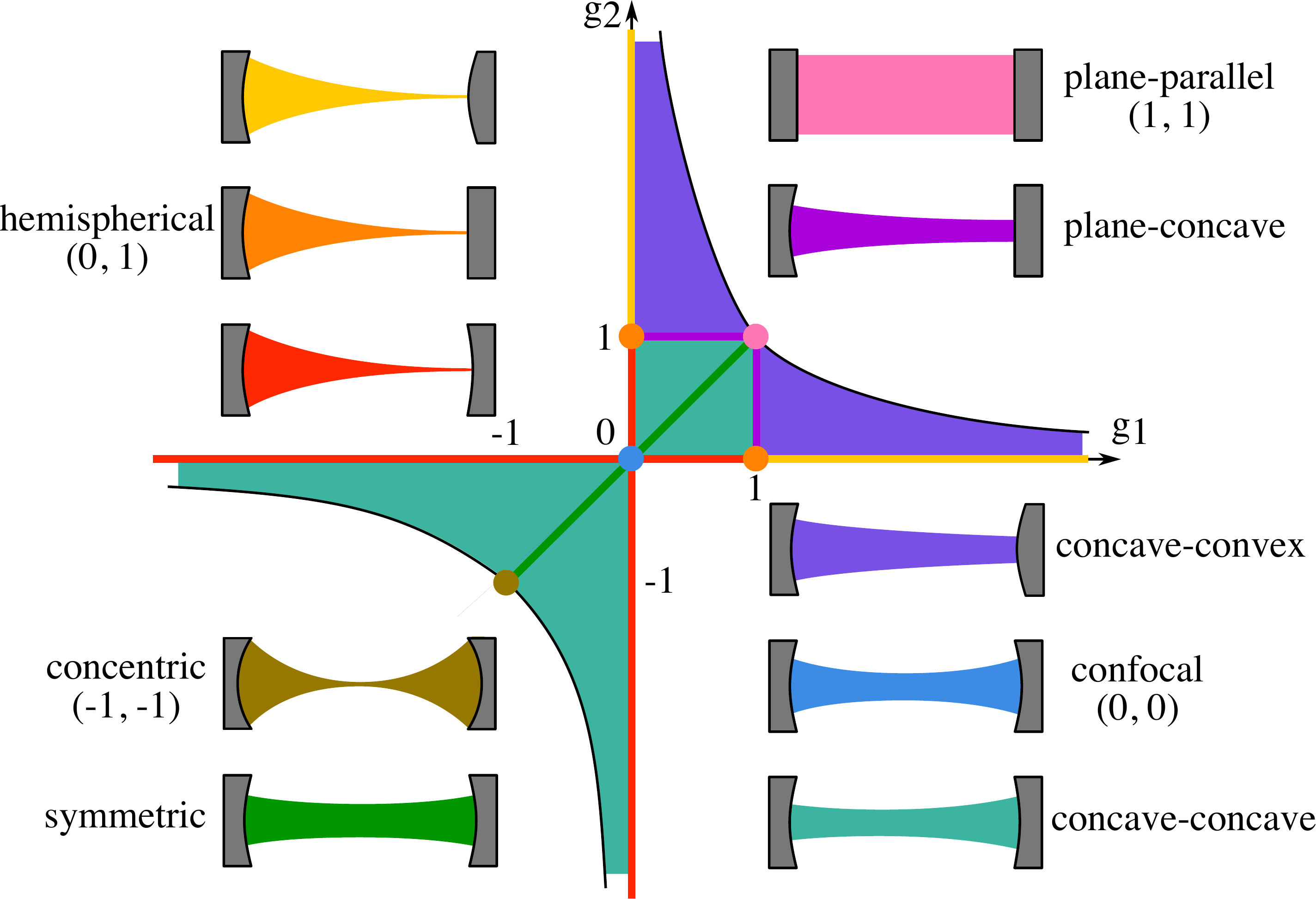}
\caption{Stability diagram of a Fabry-Perot cavity. All coloured regions bounded by the curves $g_{1,2}=0$, and $g_{1}g_{2}=1$ correspond to stable configurations. The coloured cavity configurations correspond to the regions of the diagram with matching colour. Cavities close to the edges are less stable.}
\label{fig:optical_cavity}
\end{center}
\end{figure}

The geometrical stability of an optical cavity is parametrized by its so-called $g$-factor. The $g$-factor can be thought of as the magnification experienced by an incident collimated beam upon reflecting once from each mirror, and returning to the input mirror. For a Fabry-Perot cavity the $g$-factor can be shown to be~\cite{siegman1986lasers}:
\begin{equation}
g_{c} = g_1 g_2 = \left(1-\frac{L}{R_{1}}\right)\left(1-\frac{L}{R_{2}}\right),
\label{equ:g-factor}
\end{equation}
where $g_{1,2}=1-L/R_{1,2}$ are the $g$-factors of the mirrors, $L$ is the cavity length, and $R_{1,2}$ are the radii of curvature (RoC) of the mirrors. The sign criterion for the RoC is that it is positive for mirrors concave towards the cavity centre and negative otherwise. Geometrical stability is a purely structural property of the cavity, depending only on the curvature of the mirrors and the cavity length (and not, for instance, on the reflectivities of the mirrors). If a beam would maintain a finite size given an unlimited number of reflections by the cavity, the cavity is said to be stable. If the beam spot eventually grows without limit under repeated reflection within the cavity, it is said to be unstable. The condition for cavity stability translates into:
\begin{equation}
	0\leq g_{c}\leq1.
\end{equation}

Fig.~\ref{fig:optical_cavity} shows the stability diagram of the cavity as parametrized by $g_1$ and $g_2$. Current gravitational wave detectors use arm-cavities that have large beam spots on both mirrors. This is achieved by pushing the cavities into the concentric regime, while remaining stable ($g_{c} \approx 0.8-0.9$). Our tabletop cavity is plane-concave and is pushed towards the hemispherical limit. In a hemispherical cavity the beam spot is large on the curved mirror and small on the plane mirror; it is equivalent in many ways to a concentric cavity of twice the length. For many applications cavities will be designed to lie well within the stable region, as this reduces the sensitivity to changes in cavity length and alignment of the mirrors, and to thermally induced fluctuations of the mirror shape. There are, however, a few situations in which a near-unstable configuration offers an advantage. For instance, in some experiments, a relatively large or small mode volume may be desired~\cite{Bertet2001, durak14}. An important application of NUCs is in gravitational wave detectors. Using NUCs for the arm cavities of laser interferometers such as LIGO~\cite{AdvancedLIGO15} would result in larger beam spots on the test masses. The beam would then sample a larger surface area on the mirror, averaging over more of the thermal fluctuations in the mirror surface, thus improving the sensitivity of the detector to mirror coating thermal noise. Reducing the thermal noise by changing the intensity distribution of the beam can also be achieved using alternative beam shapes, such as Laguerre-Gauss modes~\cite{Mours06, Bond11, Sorazu13}. Alternative beam shapes present unique difficulties that make them challenging to implement in gravitational wave detectors. For an overview of alternative beam shapes in interferometry see  chapter 13 `Beam shaping' in~\cite{harry2012book}.

NUCs have already been employed for the recycling cavities of the Advanced Virgo detector~\cite{AdvancedVirgo15}, where they have proven to be very challenging for alignment and thermal control~\cite{Rocchi11}. NUCs with suspended mirrors are to be investigated further for the AEI 10\,m prototype experiment~\cite{gossler10}. However, it may also prove difficult to maintain length and alignment control of a suspended cavity to the precision required in the near-unstable regime~\cite{sidles06}. Despite the challenging work ahead, the resulting reduction of coating thermal noise is worth the effort, and justified by the need of a comprehensive study of the optical properties and behaviour of NUCs, we have designed a tabletop NUC experiment. This experiment constitutes a fast and powerful tool from which to extrapolate solutions for future detectors.

This paper is organized as follows: In Sec.~II we describe the design challenges and choices associated with near-unstable cavities, as well as the principle of operation of the tabletop experiment and the cavity parameters used. In Sec.~III we detail the optical setup of the cavity and the control systems employed to allow precise frequency measurements. In Sec.~IV we present the main results of the experiment, quantify the geometrical stability of the cavity as it is ``pushed beyond the edge'' of stability, and use a simple model of the cavity to give an estimate of mirror astigmatism. Lastly, in Sec.~V we discuss the implications of these results for future gravitational wave detectors.

\section{Optical Design}\label{sec:optical design}

\subsection{Design considerations for near-unstable cavities}

\begin{figure}
\begin{center}
\includegraphics[width=0.99\linewidth]{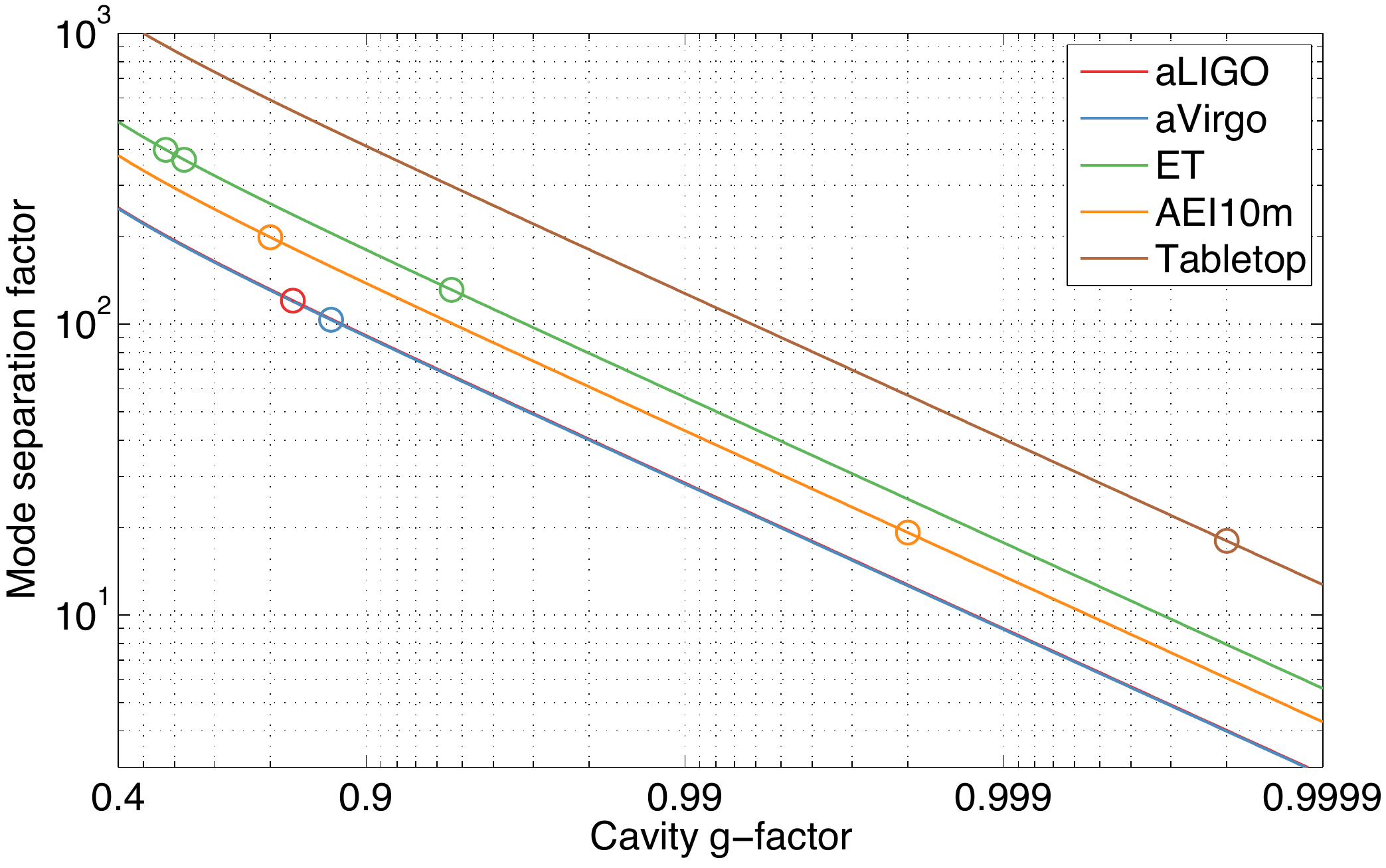}
\caption{The separation factor of the first order mode $\delta_1$ as a function of the cavity $g$-factor in different gravitational wave detectors. The dots represent the separation factors in their current configurations. They all use mode separation factors above 100, except from the AEI 10\,m prototype with $\delta_1 \approx 20$ when pushed to its maximum design stability. Our tabletop cavity, with a finesse of around 2000, has separation factor similar to the prototype at its targeted stability.}
\label{fig:sep_factor}
\end{center}
\end{figure}

One of the major difficulties of working with NUCs is that higher order modes may become co-resonant with the fundamental mode. The phase shift accumulated by higher order modes relative to the fundamental mode after a round-trip in the cavity is referred to as the round-trip Gouy phase. In Fabry-Perot cavities this is given by~\cite{RTGouy}:
\begin{equation}
\Delta \Psi_{N} = N \cdot 2 \arccos \pm \sqrt{g_c},
\end{equation}
where $N$ is the order of the mode and the sign $\pm$ is determined by the sign of $g_1$~\cite{RTGouy}. Thus, the offset in resonance frequency of the $N$-th order mode with respect to the fundamental mode is given by:
\begin{equation}
f_{N}= \frac{\Delta \Psi_{N}}{2\pi} \FSR  = N\cdot \text{FSR}\cdot \frac{\arccos \pm \sqrt{g_{c}}}{\pi}.
\label{equ:MSF}
\end{equation}
where $\text{FSR} = c/(2L)$ is the free spectral range of the cavity. For a near-concentric cavity the round-trip Gouy phase of the $N$-th order spatial mode is close to $N \cdot 2\pi$, and in consequence its offset in resonance frequency is close to $N \cdot \text{FSR}$ (all higher order modes are co-resonant with the fundamental mode). A near-hemispherical cavity has half the round-trip Gouy phase of a near-concentric, and hence the $N$-th order mode is offset by $N \cdot \text{FSR}/2$ (even higher order modes are co-resonant).

In order to quantify the separation in resonance frequency independently of the cavity linewidth, it is useful to introduce the \textit{separation factor}:
\begin{equation}
\delta_{N}=\frac{\Delta f_{N}}{\text{FWHM}/2},
\label{equ:sep_factor}
\end{equation}
where $\Delta f_{N}$ is the frequency offset between the $N$-th order mode resonance and the nearest fundamental mode resonance (see Fig.~\ref{fig:PC_MSF_t}), $\text{FWHM} = \text{FSR}/\finesse$ is the linewidth of the cavity, and $\finesse$ is the finesse. More details of the definition of $\Delta f_{N}$ can be found in~\cite{Thesis}.

Current gravitational wave detectors use configurations close to the concentric but well within the stable region. For instance, the Advanced LIGO detectors~\cite{AdvancedLIGO15} employ arm cavities with $g$-factors of 0.832. Fig.~\ref{fig:sep_factor} shows the separation factor as a function of cavity stability in Advanced LIGO (aLIGO), Advanced Virgo (aVirgo), the Einstein Telescope (ET)~\cite{ETconceptual2011, hild2008push}, the AEI 10\,m prototype and the tabletop cavity presented in this paper.

The separation factor is a useful parameter for understanding the limitations of the cavity in terms of geometrical stability, i.e., it provides a good indication of the upper limit of the cavity $g$-factor in order to avoid mode degeneracies. In practice, the separation factor should be substantially greater than $1$ so that the cavity has some tolerance to mirror imperfections and alignment errors. There is a tradeoff between $g_c$ and $\mathscr{F}$: at low $g_c$ a high finesse is necessary in order to achieve a high enough separation factor.

\subsection{The tabletop experiment}

The optical setup presented in this paper allows the study of the behaviour of a cavity as it is parametrically pushed close to the boundary of geometrical stability and beyond. We set up a half-symmetric plane-concave cavity which we push into the near-unstable and unstable hemispherical configuration. This design was chosen over the symmetric concentric cavity for ease of mode matching and length stabilization.

The cavity is formed by a flat input mirror (IM) and a concave end mirror (EM). The RoC of the EM was chosen to be 1\,m so that the cavity length in the near-unstable regime is approximately 1\,m. In the near-unstable regime $g_{c}\rightarrow 0$ with $g_{1}=1$ and $g_{2}\rightarrow 0$.

The plane-concave cavity can be viewed in many ways as a concentric cavity of twice the length, but there is a critical difference between them due to their different cavity lengths. In a near-hemispherical cavity the accumulated round-trip Gouy phase is half that of the near-concentric. Thus, only even-order modes become co-resonant with the fundamental mode in near-instability, with odd-order modes becoming anti-resonant. As we will show, when close to the boundary of geometrical instability, the cavity's resonance conditions become highly sensitive to mirror surface imperfections, which leads to changes in the mode spacing and in the spatial profile of the beam.

The $g$-factor of our tabletop cavity can be changed easily and our aim is to investigate a regime that is at least as close to instability as the proposed arm cavities of the AEI 10m prototype. A high enough finesse is then required to allow sufficient mode separation. We choose a finesse of approximately 2000 to obtain a mode separation factor of around 20 for our targeted maximum $g_c$, which allows us to acquire a good mode separation while keeping a simple length sensing and control scheme, and offers a similar separation factor to the AEI 10m prototype arm cavities at its targeted maximum $g$-factor of 0.998.

Table~\ref{tab:parameters of PCC} lists the design parameters of the tabletop cavity. The cavity is first set up to lie well within the stable region, with a $g$-factor similar to that of the arm cavities in Advanced LIGO and Advanced Virgo. Once the cavity is aligned and locked, its length is incrementally adjusted in order to push it towards the near-unstable and unstable regions.

\begin{table}
\begin{ruledtabular}
\begin{tabular}{ccccc}
Cavity length (m) & 0.956 & 0.993	& 0.999 & 0.9999 \\
\hline
Beam waist ($\mu$m) & 263.56 & 168.04 & 103.46 & 58.19 \\
Beam spot at EM (mm) & 1.26 & 2.01 & 3.27 & 5.82 \\
Rayleigh range (mm) & 205.10 & 83.37 & 31.61 & 10.00 \\
Divergent angle (mrad) & 1.29 & 2.02 & 3.27 & 5.82 \\
FSR (MHz) & 156.80 & 150.95 & 150.05 & 149.91 \\
$f_{1}~(\times \text{FSR})$ & 0.433 & 0.474 & 0.490 & 0.497 \\
$f_{2}~(\times \text{FSR})$ & 0.865 & 0.947 & 0.980 & 0.994 \\
$\delta_{2}$ & 563.2 & 223.2 & 84.6 & 26.6 \\
$g_{c}$ & 0.044 & 0.007 & 0.001 & 0.0001 \\
$g_{c}^{*}$ & 0.832 & 0.972 & 0.996 & 0.9996 \\
\end{tabular}
\end{ruledtabular}
\caption{Parameters of the tabletop cavity at four different values of the cavity length. $g_c$ is the corresponding cavity $g$-factor, and $g_{c}^{*}$ is that of the equivalent near-concentric cavity that has the same eigenmodes but twice the length.}
\label{tab:parameters of PCC}
\end{table}

\begin{figure}[!ht]
\begin{center}
\includegraphics[width=0.99\linewidth]{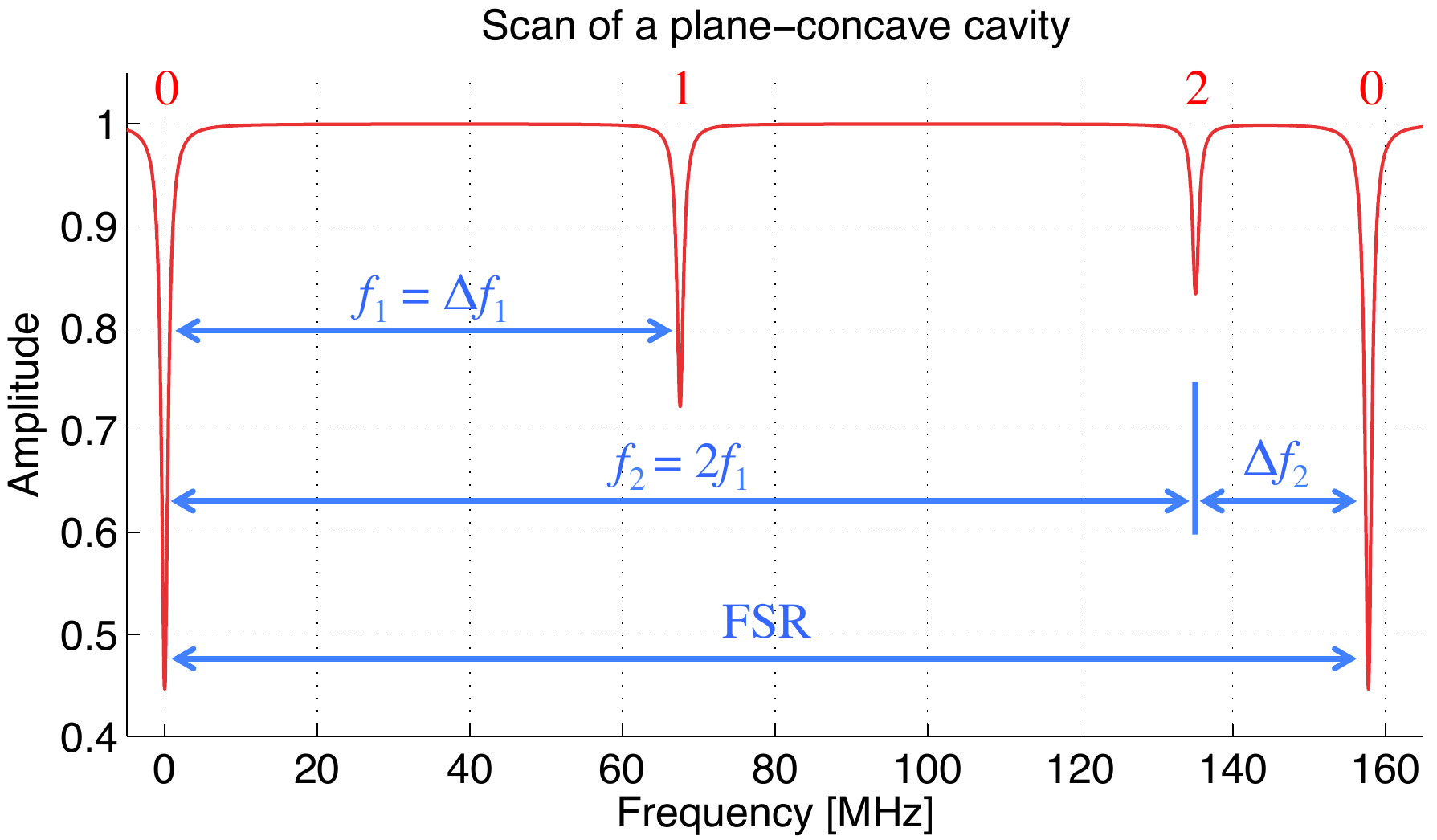}
\caption{Simulation showing the resonant frequencies of the fundamental mode and the first and second order spatial modes in the tabletop cavity in its initial stable configuration. Offsets in resonant frequencies can be measured and used to extract information about the cavity stability. The resonances of all modes have periodicity over the cavity's free spectral range. In the near-unstable plane-concave cavity the even-order modes become co-resonant with the fundamental mode, and the odd order modes become anti-resonant.}
\label{fig:PC_MSF_t}
\end{center}
\end{figure}

In order to precisely characterize the stability of the cavity, we measure the resonance frequency of the second order mode $f_{2}$ (see Fig.~\ref{fig:PC_MSF_t}), and derive expressions for $g_2$ and $L$ in terms of $f_{2}$. In a plane-concave cavity the round-trip Gouy phase is alternatively written as:
\begin{equation}
	\Delta \Psi_N = N \cdot 2 \arctan\left( \frac{L}{z_R}\right),
\end{equation}
thus, using using Eq.~\ref{equ:MSF} for $N=2$:
\begin{equation}
f_{2} = 2\frac{\text{FSR}}{\pi} \arctan\frac{L}{z_{R}},
\label{equ:MSF2}
\end{equation}
where $z_{R}$ is the Rayleigh range, which for a plane-concave cavity is given by:
\begin{equation}
	z_{R}=L\sqrt{\frac{g_2}{1-g_2}}.
\end{equation}
This yields the following expressions:
\begin{equation}
g_2 = \frac{1}{2} \left[ 1+\cos \left( \frac{f_{2}}{\text{FSR}} \pi \right) \right],
\label{equ:g_vs_df}
\end{equation}
and
\begin{equation}
L = R_{2} \left( 1- g_2 \right) = \frac{R_{2}}{2} \left[ 1-\cos \left( \frac{f_{2}}{\text{FSR}} \pi \right) \right].
\label{equ:L_vs_df}
\end{equation}
The resonance frequency $f_2$ thus becomes an accurate probe of the geometrical stability of the cavity. By making precise frequency measurements of $f_2$ and the FSR, we can use Eq.~\ref{equ:L_vs_df} to accurately determine the RoC of the mirror. It will turn out, as we detail in Sec.~IV, that the second order mode resonates at slightly different frequencies depending on the orientation of its spatial profile. We will use this fact along with Eq.~\ref{equ:L_vs_df} to give an estimate of the mirror astigmatism.

\section{Optical Setup}

\begin{figure*}[!ht]
\begin{center}
\includegraphics[width=0.99\linewidth]{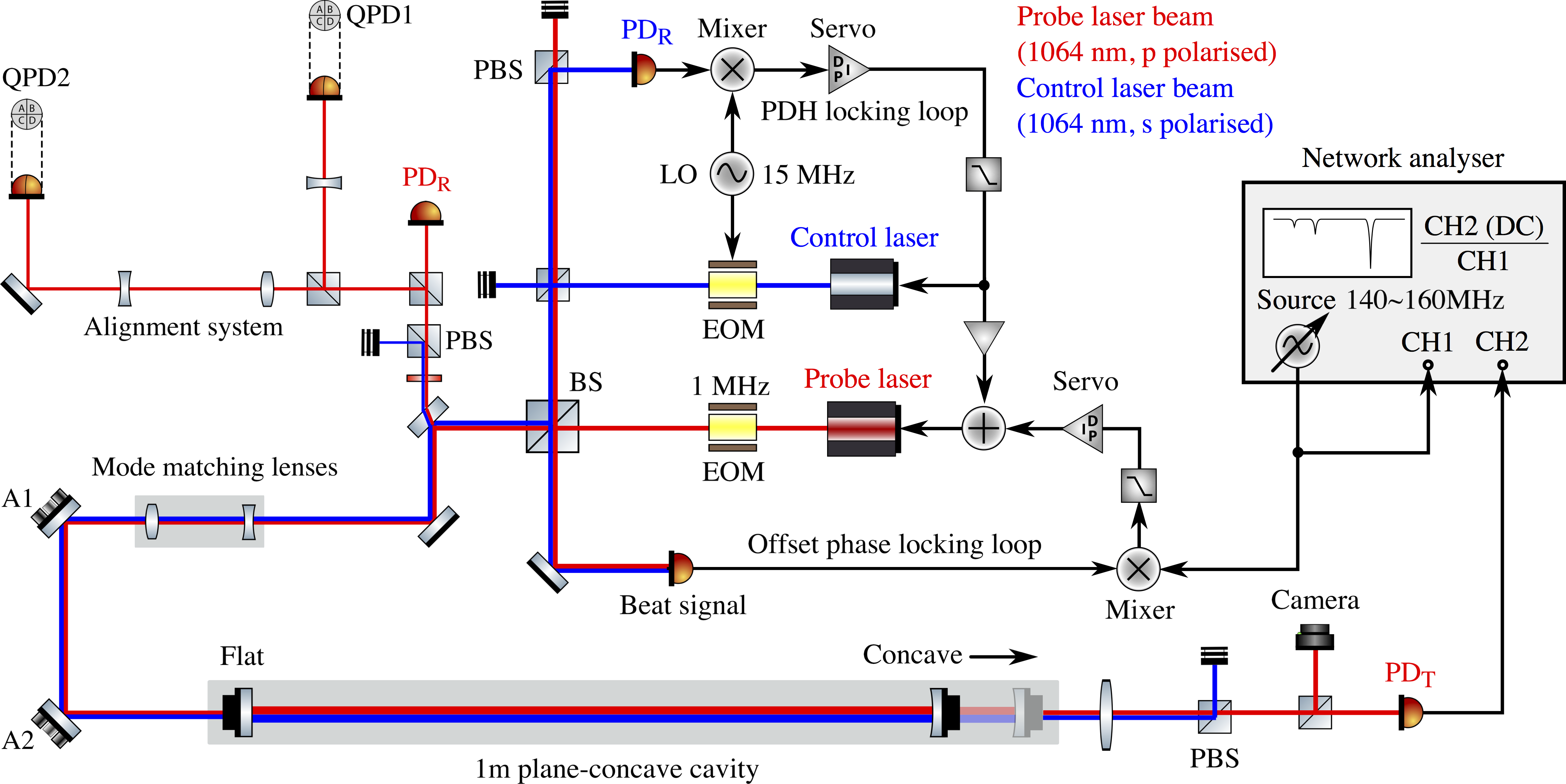}
\caption{Schematic of the near-unstable cavity setup. The blue and red lines represent the beams from the 1064\,nm control and probe lasers respectively. They pass through phase modulators (EOM) that generate the sidebands used for control purposes. The beams are combined at the central beam splitter (BS) and then coupled into the cavity through two mode matching lenses. The beams have orthogonal polarizations upon combination to ensure that they don't interfere. The control laser is locked to the cavity via a Pound-Drever-Hall loop (PDH). A fraction of the PDH signal is fed back to the probe laser to reduce common mode frequency noise. The probe laser is then locked to the control laser with a tunable frequency offset via an offset phase locking loop. The frequency offset is controlled by the source of a network analyser. The probe laser is then used to study the cavity. A camera and a photodetector (PD) are placed at the cavity's transmission port to observe the resonant mode. The subscripts of R and T stand for reflection and transmission ports of the cavity, respectively. An alignment sensing system based on wavefront sensing is used to help maintain accurate alignment of the cavity.}
\label{fig:setup_full}
\end{center}
\end{figure*}

The optical layout is presented in Fig.~\ref{fig:setup_full}. Two commercial infrared lasers are used: the ``control laser'' (Mephisto 300 NE), and the ``probe laser'' (Mephisto 1200 NE) from Innolight, Germany. Both are diode-pumped single frequency continuous-wave solid-state lasers, using a monolithic Nd:YAG crystal in the non-planar ring oscillator configuration~\cite{KB85}. They provide an output frequency tunable over 30\,GHz, centred at a wavelength of 1064\,nm, and have a very narrow linewidth of 1\,kHz over 100\,ms. The optical setup that immediately follows each laser is identical (not shown), with the beam passing through a Faraday isolator to prevent back-reflections into the laser, polarizing optics and mode-matching telescopes. The two beams are combined at the central beam splitter, at which point they have orthogonal polarization to ensure that no interference occurs. The beams are then coupled into the cavity via the flat input mirror. Before being combined, each beam passes through an electro-optic modulator (Newport Corporation Model 4004) that generates phase modulation sidebands in radio frequencies for control purposes. The control laser is modulated at 15\,MHz to provide the error signal for a Pound-Drever-Hall (PDH) loop~\cite{drever83}. The control laser is locked to the cavity by actuating on its piezo-drive, and the probe laser is in turn locked to the control laser with a tunable frequency offset. The probe laser can then be used to study and characterize the cavity, e.g., to quantify its geometrical stability via measurements of the FSR and the resonant frequencies of higher order modes.

\subsection{Mode matching group}\label{sec:mode matching}

Both lasers have their own mode matching telescopes before combination at the beam splitter. This ensures that both the control and the probe beams have the same parameters after combination, yielding a beam waist of 371\,$\mu$m. Two lenses with focal lengths of $\pm200$\,mm are then used to mode match the combined beams to the cavity. These lenses are installed on a rail so that their positions can be set manually to adjust the mode matching to different cavity configurations. The lenses are carefully aligned so that the beams go through the centre to a high precision, in order to avoid the effect that the longitudinal motion of the lenses can have on the transverse position of the beam spot in the input mirror.

The space available for moving this lens group would, in principle, allow for mode matching to the cavity as its geometrical stability is varied from $g_{c}=0.044$ (with $w_{0} = 263.56$\,$\mu$m) to $g_{c}=0.00004$ (with $w_{0} = 46.28$\,$\mu$m). In practice, however, there are two factors limiting the precision of mode matching: (1) the difficulty to measure precisely the size and position of the resulting beam waist, and (2) the precision to which the position of the lenses can be adjusted. We find that it is best to first use the position and size of the beam waist to carry out a rough mode matching, and then use the cavity itself for fine tuning. As mode mismatch translates into coupling of the fundamental mode to the second order spatial mode, we optimize mode matching by adjusting the positions of the two lenses until the reflected peak of the second order mode resonance is minimized. Since the second order mode content depends non-linearly on the lens positions, it is not always easy to find their optimal positions.

\subsection{PDH stabilization system}\label{sec:PDH stabilisation}

An essential requirement in order to obtain reliable measurements of mode spacing is that the laser must remain stabilized and aligned to the cavity. The frequency stabilization is carried out using the Pound-Drever-Hall (PDH) loop with the 15\,MHz sidebands of the control laser. The reflected control beam is captured in a photodetector, whose signal is demodulated with a 7\,dBm mixer circuit (ZAD-1-1+ from MiniCircuits) to produce a linear error signal. This is then fed to a digital servo system produced in house, which drives the piezo actuator on the control laser via a high-voltage amplifier. This servo can provide a gain of up to 10. It incorporates three integrators with cut-off frequencies at 100\,Hz, 1\,kHz, and 10\,kHz respectively to introduce extra gain below these frequencies. The digital servo also automates the initial frequency scan and lock acquisition.

\subsection{Offset phase locked laser}\label{sec:offset phase locking}

The probe laser is phase-locked to the control laser via a two-stage system: First, a fraction of the PDH signal from the control laser is fed to the piezo actuator of the probe laser, which greatly reduces all common mode frequency noise due, e.g., to air turbulence in the cavity, thermally induced motions or acoustic resonances of the mirrors. The PDH loops for the two lasers will in general have different gains, so the gain of the probe laser loop can be adjusted with a variable attenuator to minimise this difference. In the second stage, the polarization of the two beams is adjusted so that they interfere on a photodetector, producing a beat signal of frequency equal to their frequency difference. This signal is fed to a mixer, where it is combined with a local oscillator (LO) signal, producing an error signal proportional to the difference between the frequency offset and the reference signal, which is used to maintain this offset frequency at a target value.
 
Once the offset lock is acquired, the frequency of the local oscillator can be tuned to adjust the frequency difference between the two lasers over several FSR. This allows us to scan the cavity with the probe laser while using the control laser to maintain cavity lock. This is crucial since, as the cavity gets close to the edge of stability, small changes in the cavity length or alignment have a large impact on the cavity's eigenmodes.

\subsection{Alignment sensing system}

The alignment of the cavity relative to the input beams is sensed using the wavefront sensing technique proposed by Anderson~\cite{anderson84, Morrison94a}. We use a 1\,MHz sideband of the probe laser produced by an EOM, and rather than measuring the reflected modulated signal in the fundamental mode, we measure the higher-order mode content of the reflected beam.

Light resonant in the cavity will interfere with the incoming beam if the cavity is slightly misaligned, and produce coupling of the fundamental mode to higher order modes in the reflected beam. For small misalignments, the higher order mode components provide signals proportional to the cavity displacement and tilt angle. Translation of the cavity axis or the input beam axis produces light in the first order mode, and the coupling factor is proportional to the amplitude of the translation. The tilt of the cavity axis also couples the fundamental mode to the first order mode, with a phase shift of 90$^{\circ}$ relative to the original beam. The reflected beams are sent to two quadrant photodiodes (QPD), which make a differential measurement of the beam profile. In order to separate the angular signal from the displacement signal, we induce a Gouy phase shift of 90$^{\circ}$ between the two detection ports.

The layout of the alignment system is shown in Fig.~\ref{fig:setup_full}. The beam reflected by the cavity propagates back through the mode matching lens group and is sent to the QPDs. This makes the alignment sensing system independent of the mode matching system and the cavity configuration. The signals from the QPDs are output through differential amplifier circuits and demodulated with a pair of mixers using the same local oscillator signal used to modulate the probe laser at 1\,MHz.

The 4 DC signals output from these mixers are linear signals of the angular and transverse displacement errors of the cavity relative to the input beams. The error signals are not fed back to the system automatically but provide an accurate guide for our manual alignment. When this sensing system is engaged along with the PDH lock, the input beam is controlled in 5 out of its 6 degrees of freedom, all but the polarization direction, which is not expected to significantly fluctuate.
More details about the experimental setup used in this work, especially the offset phase locking loop, the alignment sensing systems and derivations of their error signals, are described in chapter 4 in~\cite{Thesis}.

\section{Results}

With the experimental setup presented in the previous section, we investigate the behaviour of the cavity when pushing it beyond the edge of geometrical stability. The two main purposes are: (1) to find out how the cavity eigenmodes behave in a practical setup, and (2) to determine how far we can push the cavity towards the edge of geometrical instability before it becomes no longer of practical use, due e.g. to high optical loss or mode degeneracy.

\subsection{Influence of mirror surface imperfections}

\begin{figure}[!ht]
\begin{center}
\includegraphics[width=0.99\linewidth]{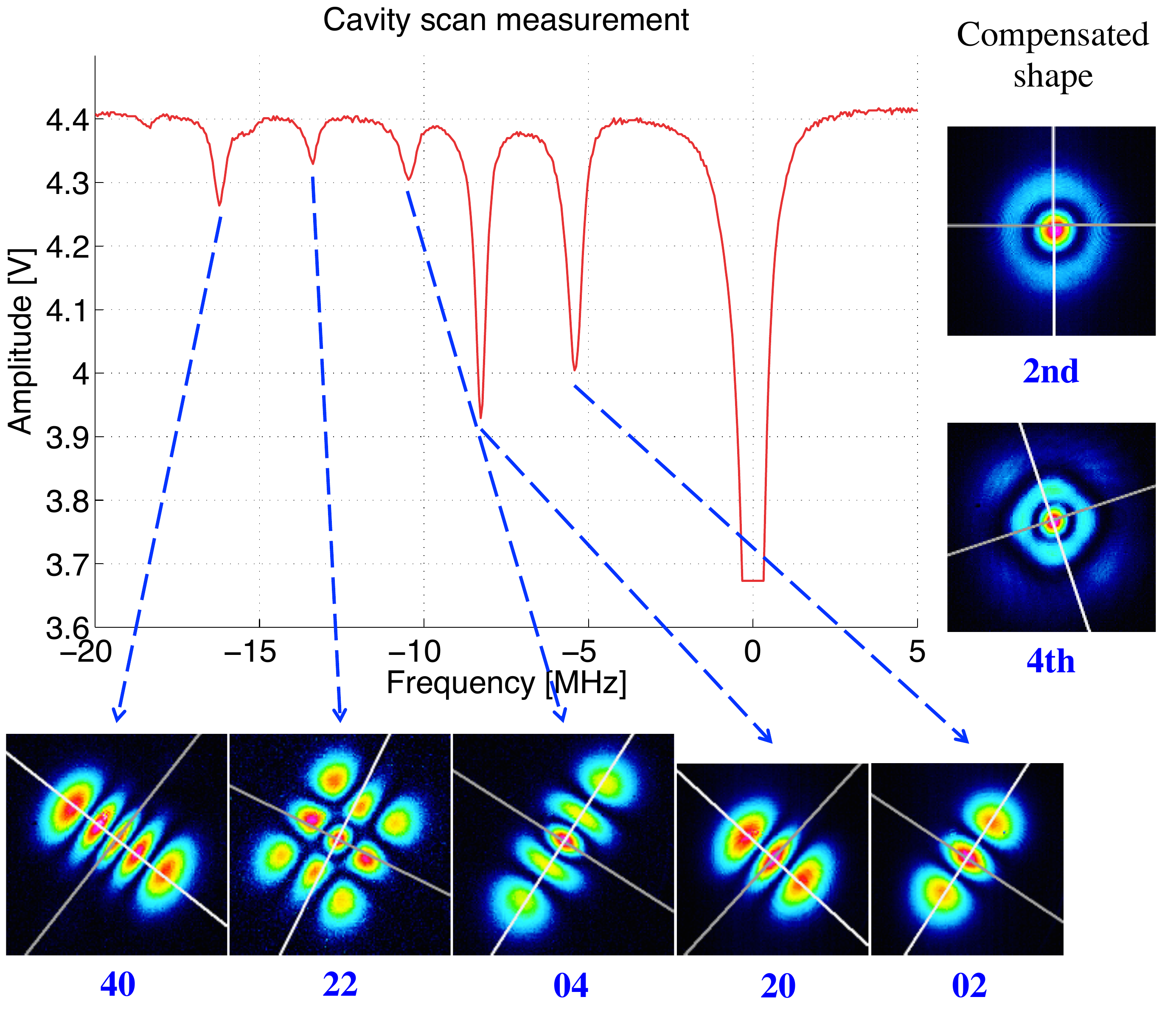}
\caption{A cavity scan measurement showing the resonant frequencies the fundamental mode, and the second and fourth order spatial modes. The resonances of modes of the same order but orthogonal spatial profile appear separate due to the imperfect surface of the spherical mirror. The shapes of the modes as they are scanned are shown in the bottom. The separation can be reduced by increasing the stress of the screw holding the spherical mirror, thus compensating the surface deformation. The shapes of the compensated second and fourth order modes are shown in the right, which appear as circularly symmetric Laguerre-Gauss modes.
\label{fig:ModeShape_0509_L1_t}}
\end{center}
\end{figure}

The cavity is pushed to the edge of geometrical stability and beyond ($g_c<0.001$). The Hermite-Gauss eigenmodes (\HG{nm}) resonant in the cavity are observed, and degenerate modes of the same order appear clearly separated in resonance frequency (see Fig.~\ref{fig:ModeShape_0509_L1_t}). Normally, resonances of the same order should overlap in frequency because they experience the same Gouy phase shift in a cavity round-trip. The fact that these modes appear separate in resonance frequency is a good indication that the mirror surface deviates from being perfectly spherical. The interaction of the circulating field in the cavity with the imperfect mirror surface results in modes of the same order but orthogonal spatial profile accumulating a different phase as they propagate from mirror to mirror and back.

It was observed that the modes had their profile oriented at an angle of 45$^{\circ}$ with the horizontal, and the separation in resonance frequency could be compensated by tightening the screw holding the mirror in the mirror mount, which was also at 45$^{\circ}$ with the horizontal. Fig.~\ref{fig:ModeShape_0509_L1_t} shows the shapes of the second and fourth order modes before and after tightening the screw. The resonance peaks that appear separate due to the imperfect mirror could be made to overlap by applying structural stress to the mirror in a way that compensated for its default deformation, leading to the circularly symmetric shapes shown similar to Laguerre-Gauss modes.

The effect of the imperfect mirrors on the resonances of degenerate spatial modes is usually negligible in geometrically stable cavities, and becomes significant as the cavity approaches near-instability. Mirror surface deformations become a dominant factor in determining the behaviour of the spatial modes when the cavity is close to the boundary of geometrical stability.

\subsection{Quantifying cavity stability}

The end mirror is mounted on a translation stage (Newport Corporation M-460A) with a differential micrometer (DM-13) which enables us to control the position of the mirror along the optical axis with a resolution of 0.5\,$\mu$m. We rotate the end mirror by approximately 45$^{\circ}$ so that the \HG{nm} modes become aligned with the transverse axes.

From the starting position of the translation stage, with a cavity length of roughly 999\,mm, the position of the end mirror is adjusted 18 times to increase and decrease the cavity length, moving towards and away from geometrical instability. Each time the cavity length changes we repeat the following procedure: (1) rough mode matching using the beam waist size and position, (2) fine mode matching using the reflected beam, (3) fine alignment using the QPDs, and (4) PDH lock. The cavity is then scanned with the offset phase locked probe laser. The source of a network analyser, which can produce a swept sinusoidal signal, is used as the tunable oscillator. The reflected light is detected, revealing the resonance structure of the cavity, and the transmitted beam is observed with a camera in order to characterise its spatial properties.

\begin{figure}[!ht]
\begin{center}
\includegraphics[width=0.99\linewidth]{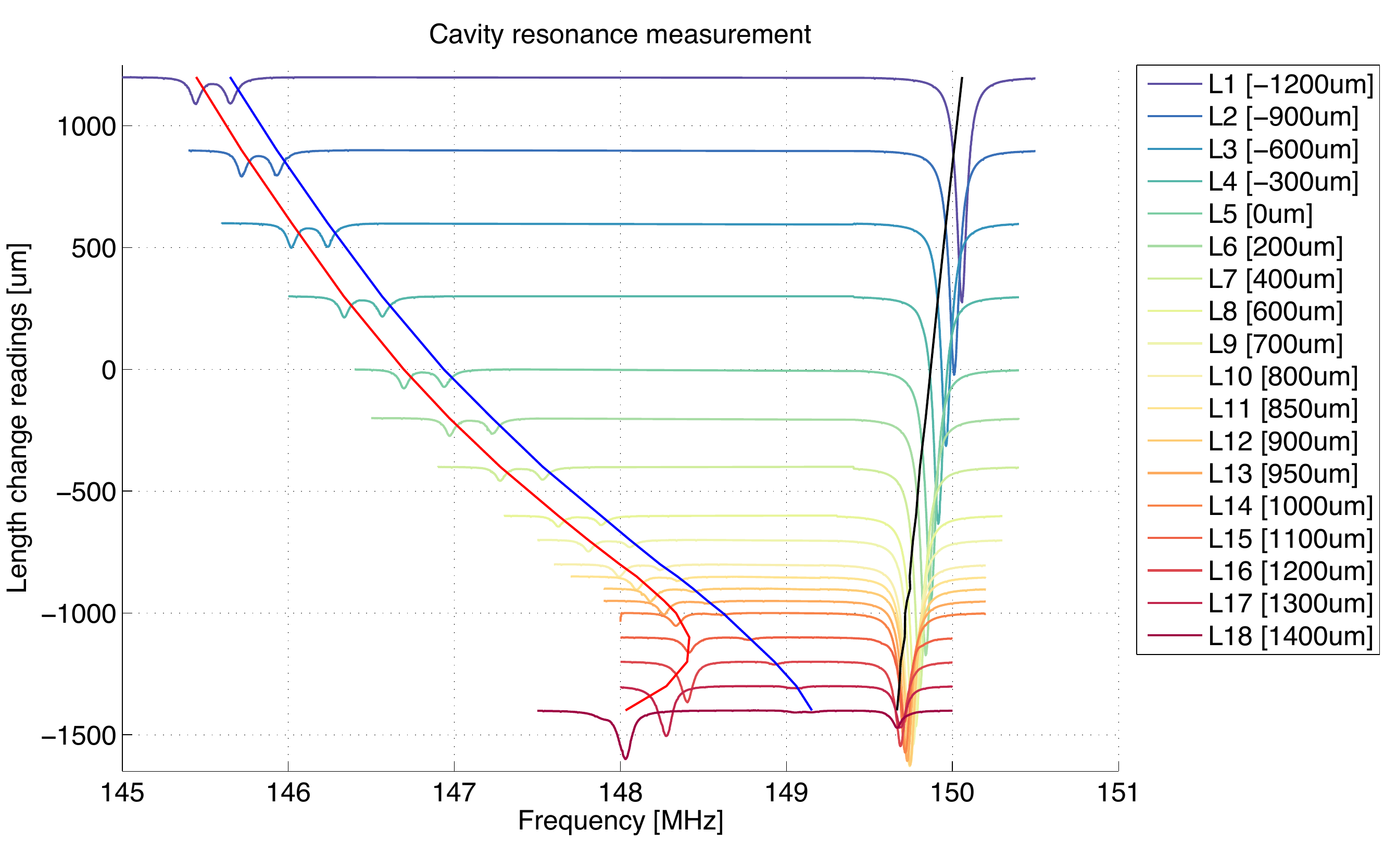}
\caption{Resonances measured by scanning the frequency of the probe laser when pushing the cavity to near-unstable region. We track resonant frequencies of the fundamental mode, the HG02 and HG20 modes, which are plotted in black, blue and red respectively. L5 is the position where the translation stage gives a reference reading of roughly 999\,mm. Legends from L1 to L18 display relative length changes from this value. The y-axis represents the cavity length deviation from L5 for each measurement. Peaks on the right are resonant frequencies of the fundamental mode, which give a very good measurement of the free spectral range of the cavity. The two peaks on the left at first represent resonances of the separated HG20 and HG02 modes. Mode separation frequencies between second order modes and the fundamental mode become smaller and smaller. But beyond a certain point, it seems that second order modes begin to move back away from the fundamental mode. We believe that what we tracked are no longer the pure second order modes; indeed, the measured beam profiles appear to have an increasing fraction of higher order mode content. It could be that the cavity has in these cases become unstable, but these modes are near-resonant in the cavity.
\label{fig:peaks_change}}
\end{center}
\end{figure}

We measure the separation frequency between the second order mode and the fundamental mode, $f_2$, and from it derive the length and stability of the cavity using Eqs.~\ref{equ:g_vs_df} and~\ref{equ:L_vs_df}. Fig.~\ref{fig:peaks_change} presents a series of cavity scans showing the resonance peaks of the fundamental mode and the second order mode. Each scan consists of 800 points in a range of 2.5\,MHz, resulting in a scanning resolution of 3.125\,kHz. The cavity length is adjusted from its initial position, $L_0$, from $L_0-$1200.0\,$\mu$m (L1) to $L_0+$1400.0\,$\mu$m (L18).

From L1 to L7, the amplitude of the first order mode can be minimized to almost to zero during mode matching and alignment. After L7, however, the residual first order mode increases significantly and can no longer be eliminated. This is thought to be due to the large beam suffering from clipping loss at the end mirror. The reflected beam seen by the alignment system is no longer axially symmetric, thus the difference signal between the two halves of the beam cannot be zero. Mode matching suffers a similar but more severe problem: From L1 to L18, the amplitudes of \HG{02} and \HG{20} cannot be simultaneously reduced. This is a result of the deformation of the end mirror surface. For this measurement, since we are only interested in the frequency separation and not on the amplitudes of \HG{02} and \HG{20}, the mode matching lens group was left at the optimum configuration for $g_2=0.00004$, and was not adjusted throughout the measurement. This does not affect the frequency spacings, and is responsible for the varying amplitudes of \HG{02} and \HG{20} seen in Fig.~\ref{fig:peaks_change}.

The peak to the right in Fig.~\ref{fig:peaks_change} corresponds to the resonance of the fundamental mode, \HG{00}, and is used to measure the FSR. We verify that the FSR changes linearly as a function of $L$, as expected. The two peaks on the left correspond to the resonances of the \HG{20} and \HG{02} modes, separated by the mirror astigmatism. The frequency separation between \HG{00} and \HG{20,02} is reduced as the cavity length increases ---as expected--- up to a certain point, at the edge of geometrical stability, after which the resonant peaks seem to change direction and the separation grows. The images of the transmitted beam reveal a change in the spatial structure of the beam as the cavity becomes unstable (Fig.~\ref{fig:ModeShape}).

We perform fits of the cavity length to the measured $f_2$ according to Eq.~\ref{equ:L_vs_df}:
\begin{align}
L_{0}+\Delta L &=\frac{R_{2+}}{2} \left[ 1-\cos \left( \frac{f_{20}}{\text{\,FSR}}\pi \right) \right], \nonumber \\
L_{0}+\Delta L &=\frac{R_{2-}}{2} \left[ 1-\cos \left( \frac{f_{02}}{\text{\,FSR}}\pi \right) \right],
\label{equ:L_vs_df2_fit}
\end{align}
where $f_{20}$ and $f_{02}$ are the frequency offsets of \HG{20} and \HG{02} respectively, $L_0$ is the cavity length when the translation stage is at the zero mark, $\Delta L$ is the translation stage measurement, FSR is measured for every point, and $R_{2+}$ and $R_{2-}$ are the RoCs of the end mirror as experienced individually by the orthogonal modes. This model assumes that the input mirror is perfectly flat and that the end mirror is astigmatic. The end mirror has a concave surface with a RoC of around 1\,m, but due to the surface imperfections the two orthogonal spatial modes will experience a slightly different curvature. By performing the two fits we can determine the difference in curvature.

\begin{figure}[!ht]
\begin{center}
\includegraphics[width=0.99\linewidth]{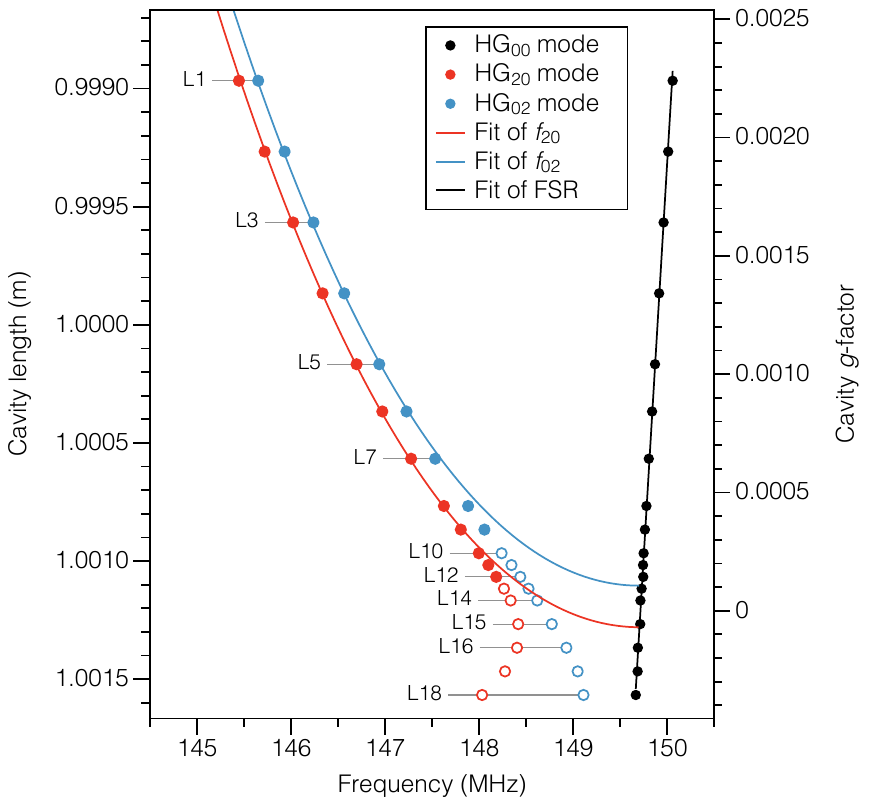}
\caption{Resonant frequencies of the \HG{00} (at one FSR), \HG{20}, and \HG{02} modes. The measurement of the FSR allows us to obtain the cavity length for every point. Fits of the cavity length to the frequencies of the \HG{20} and \HG{02} modes yield two values of the mirror RoC that differ by 145\,$\mu$m, indicating astigmatism. Only the filled points are used in the fits, as the data starts deviating from the expected behaviour for Gaussian beams as it gradually becomes unstable. The measured resonances of the \HG{20} and \HG{02} modes start to deviate from the fitting at L7 and L10 respectively where the deviation is 1\% in frequency. The cavity $g$-factor is derived as an average property by using the average RoC of the mirror. The $g$-factor suggests that the cavity becomes unstable at approximately L15, which agrees well with observations.
\label{fig:peaks_fitting}}
\end{center}
\end{figure}

\begin{figure}[!ht]
\begin{center}
\includegraphics[width=0.99\linewidth]{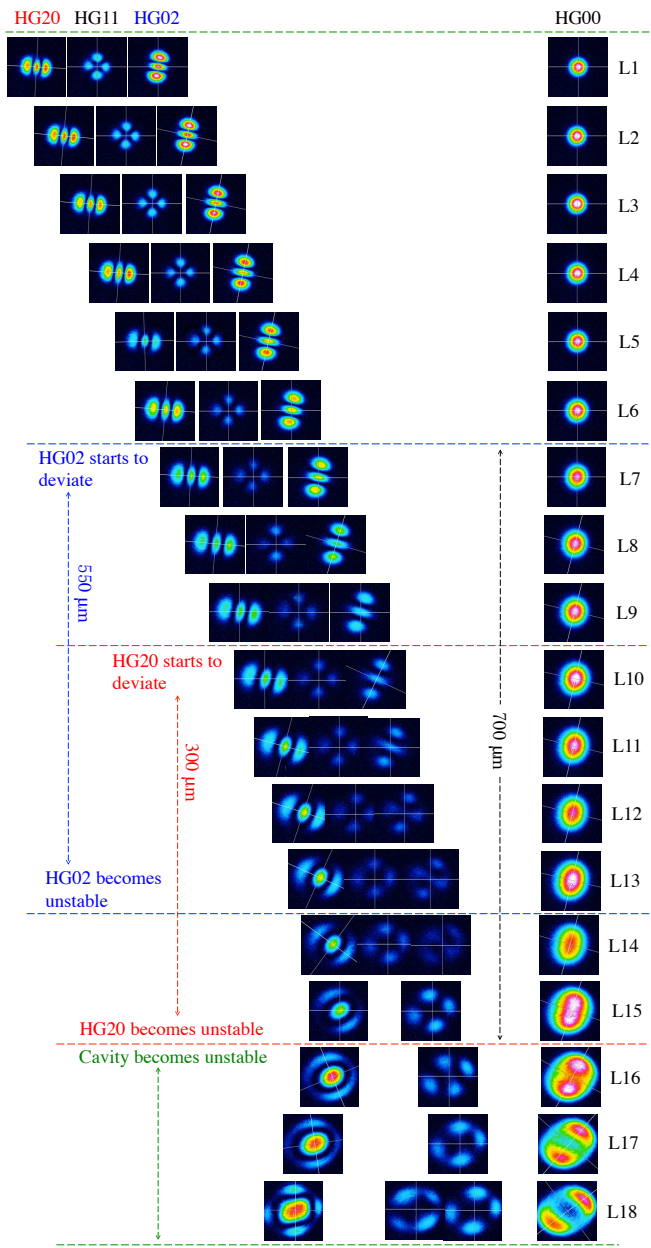}
\caption{Spatial profile of the transmitted beam at each step of the translation stage of the end mirror, the horizontal position of each image depicts the frequency of each resonance as seen in Fig.~\ref{fig:peaks_fitting}. Together with our fitting results shown in Fig.~\ref{fig:peaks_fitting}, we regard these points below as turning points. L7: \HG{02} starts to deviate from the predicted behaviour and shows a gradual clockwise rotation in its profile. L10: \HG{20} starts to deviate and shows a clear rotation. L13: \HG{02} arrives at its stability edge, however, the cavity is still stable for \HG{20}. L15: \HG{20} arrives at its stability edge and the cavity becomes completely unstable. From L13, the spatial profile of the fundamental mode goes from a typical round-spot shape to a two-spot shape, as it overlaps with the second order mode. Modes we tracked after this point no longer appear to belong to Gaussian modes if the cavity is unstable.}
\label{fig:ModeShape}
\end{center}
\end{figure}

We first calibrate the translation stage with our measurements of the FSR, and obtain $L_0=1000166.8\pm7.7\,\mu$m. We then perform the fits of Eq.~\ref{equ:L_vs_df2_fit} and obtain $R_{2+}=1001284.9\pm4.6\,\mu$m and $R_{2-}=1001140.0\pm15.7\,\mu$m. According to this model, the curvature of the mirror differs by $144.9\,\mu$m when probed individually by the second order spatial modes. We then characterize the geometrical stability of the cavity by calculating the $g$-factor using the average mirror curvature, $g=1-2L/(R_{2+}+R_{2-})$. The results are shown in Fig.~\ref{fig:peaks_fitting}.

The data agrees well with the behaviour predicted by this model up to a certain point very close to the edge of geometrical instability. The measured resonances of the \HG{20} and \HG{02} modes start to deviate from the fitting (with more than 1\% discrepancy) at L7 and L10 respectively. During each measurement for a given cavity length, the shapes of the resonant modes are recorded by the camera at the transmission port of the cavity and shown in Fig.~\ref{fig:ModeShape}. We find that the spatial properties of the transmitted beam begin to depart significantly from those expected as the cavity moves towards the geometrically unstable region.

The \HG{02} mode starts deviating from the predicted behaviour before the \HG{20} mode. This becomes more apparent when looking at the the spatial profile of the mode, that can be seen starting to change from L7. From L7 to L13, the profile of \HG{02} shows a gradual clockwise rotation. The tracked resonance is no longer that of a pure \HG{02} mode, but of a mode with mixed higher-order mode content. The resonance of \HG{02} should overlap completely in frequency with the fundamental mode at around L13 according to the model, at which point the mode would become unstable in the cavity. The resonance of the \HG{20} mode follows the predicted behaviour for a bit longer, and has started to deviate by L10. Again, from L10 to L15, we observe a rotation and deformation of its spatial profile. Between L14 and L15, the cavity is unstable for \HG{02} but still stable for \HG{20}. L15 is the edge of stability for \HG{20} in the model.

We continue tracking resonances that no longer appear to belong to pure Gaussian modes. At L13 the fundamental mode starts to suffer a spatial deformation that becomes very apparent by L15-L16: its profile goes from the typical shape of \HG{00} to a two-spot shape, as it overlaps with the second order mode. According to the average cavity $g$-factor derived from the model, the cavity is approximately hemispherical at L15.  This agrees well with our observations, and we regard the cavity as completely unstable after L15: there is high optical loss and none of the resonances appear to belong to recognizable Gaussian modes. From L7 (where \HG{02} starts to deviate) to L15 (where the cavity becomes unstable), the total cavity length change is approximately 700\,$\mu$m. Interestingly, we still obtain a reasonably clear error signal for the PDH lock and the control laser can still be stabilized to the cavity. It's worth noting that when the second order mode starts overlapping with the fundamental mode, an offset is introduced in the PDH error signal, i.e., the control laser no longer locks to a pure fundamental mode. Despite this, we are still able to obtain accurate measurements of the cavity FSR. The PDH error signals finally start degenerating dramatically when the cavity is pushed 500\,$\mu$m beyond L15.

\subsection{Further discussion}

It is found that the \HG{02} mode becomes co-resonant with the fundamental mode before the \HG{20} mode. The fact that they have different resonance frequencies indicates that the modes experience a different phase shift in a cavity round-trip, i.e., they acquire different round-trip Gouy phases. A simple model of the cavity can estimate the amount of mirror astigmatism that would cause this difference in frequencies. By fitting the data to the model, we find a RoC difference of $\approx145\,\mu$m, indicating that the mirror is slightly more curved in the vertical direction.

The separation factor between the second order mode and the fundamental mode of the plane-concave cavity is given by~\cite{Thesis}:
\begin{equation}
\delta_{2}=\frac{2\mathscr{F}}{\pi} \arccos \left( 1-2g_2 \right).
\end{equation}

In our tabletop setup, the break-down point (where the second order mode behaviour starts to deviate) happens at around $g_2=3\cdot 10^{-4}$, corresponding to a separation factor of 46. The value of the separation factor is much higher than expected. At such a point, the resonance of the fundamental mode is still far away from resonances of second order modes. It means that in our cavity with a high finesse, the mode bunching is not the critical problem that causes deviations from the mode behaviour predicted by our model. The main factor that determines the mode behaviour in this extreme near-unstable condition is instead thought to be mirror imperfections.

It is worth pointing out that mode matching was very difficult to achieve in the near-unstable region. First, small changes in the mode matching system could render the cavity unstable in an unpredictable way. Second, due to the mirror surface imperfections, perfect mode matching in which the two second order mode resonances are simultaneously minimized was impossible. A thermal control system for the end mirror could provide a way of compensating for its deformation, but it could only be applied to compensate for macro-scale imperfections of the mirror surface. Small imperfections of the surface would in the end contribute to much more complicated couplings to higher order spatial modes. These couplings become larger, more chaotic, and unpredictable in NUCs. In order to enable the use of NUCs in future gravitational wave detectors, very stringent requirements should be set on the mirror surface qualities, the length stabilization systems, and the thermal compensation systems. Mirror surface imperfections will likely be the critical factor that ultimately limits the feasibility of these devices.

\section{Summary and further work}\label{sec:conclusion}

We have constructed an optical cavity which we can use to probe an as-yet untested extreme near-unstable region, and demonstrated experimentally that the cavity eigenmode behaviour in such a cavity departs significantly from the model based on ideal mirrors and input beams. This involves a 1\,m cavity which is pushed to the edge of stability by incrementally increasing its length. We have measured a series of frequencies and shapes of resonant transverse modes, including the fundamental mode, and the separated \HG{02} and \HG{20} modes. The measured resonant frequencies were compared with our fitting results, by which the stability of the cavity is quantified.

It was found that for $g$-factors down to about $3\cdot 10^{-4}$, the cavity eigenmodes' behaviour is close to that predicted by an idealised model of a perfectly aligned cavity, with resonance spacing between the higher order modes and the fundamental mode decreasing as predicted. For smaller $g$-factors, new behaviour is observed. The resonance gap between the fundamental mode and the closest higher order modes begins to grow again, while the profile of these modes deviates from Gaussian modes such as Hermite-Gauss or Laguerre-Gauss modes. This latter effect can be understood as being due to the eigenmodes becoming much more sensitive to mirror surface distortions, which strongly breaks the cylindrical symmetry normally exhibited by the cavity. Additional modelling and simulation is required to fully understand these observations, and to determine the requirements in mirror flatness and in the angular control systems for a given $g$-factor to be feasible in a gravitational wave detector cavity. 

The work carried out in this paper motivates further experimental activity, for instance using mirrors with known significant figure errors or with thermally or otherwise deformable mirrors. It also motivates simulation work, which could eventually be used to help design and understand NUCs for gravitational wave detector cavities. The \Finesse code~\cite{FINESSE}, which is used to model cavities and related optical systems in the frequency domain, represents an ideal tool with which to carry out this program of investigation, and this work is underway within the group.

\begin{acknowledgments}
The authors would like to thank the Birmingham group including Haixing Miao, Anna Green, Daniel T\"oyr\"a, Sam Cooper and Aaron Jones for useful comments and suggestions for this project. The authors would also like to thank technical engineers David Hoyland and John Bryant who helped build electronics and software. H. Wang and A. Freise have been supported by the Science and Technology Facilities Council (STFC). M Dovale \'Alvarez and D. D. Brown acknowledge financial support from the Defence Science and Technology Laboratory (DSTL) and the UK National Quantum Technology Hub in Sensors and Metrology with EPSRC Grant No. EP/M013294/1. D. D. Brown acknowledges support from the European Commission Horizon 2020 programme under the Q-Sense project Grant No. 691156 (Q-Sense-H2020-MSCA-RISE-2015).

\end{acknowledgments}

\bibliography{mybib}

\end{document}